\documentclass[smallextended,twocolumn]{svjour3}
\usepackage{textcomp}
\usepackage{url}
\usepackage{amstext}
\usepackage{graphicx}
\usepackage[authoryear]{natbib}
\usepackage[unicode=true]{hyperref}

\usepackage{geometry}
\begin{document}

\title{Multi-scale dynamical analysis (MSDA) of sea level records versus PDO,
AMO, and NAO indexes}

\titlerunning{MSDA of sea level records vs.  PDO, AMO, and NAO indexes}

\author{Nicola Scafetta}

\institute{N. Scafetta \at
              Active Cavity Radiometer Irradiance Monitor (ACRIM) Lab, Coronado,
CA 92118, USA \& Duke University, Durham, NC 27708, USA \\
              Tel.: +1 919-225-7799\\
              \email{nicola.scafetta@gmail.com and ns2002@duke.edu}
}

\date{}

\maketitle

\begin{abstract}
Herein I propose a multi-scale dynamical analysis   to facilitate
the physical interpretation of tide gauge records. The technique uses
graphical diagrams. It is applied to six secular-long tide gauge records
representative of the world oceans: Sydney, Pacific coast of Australia;
Fremantle, Indian Ocean coast of Australia; New York City, Atlantic
coast of USA; Honolulu, U.S. state of Hawaii; San Diego, U.S. state
of California; and Venice, Mediterranean Sea, Italy. For comparison,
an equivalent analysis is applied to the Pacific Decadal Oscillation
(PDO) index and to the Atlantic Multidecadal Oscillation (AMO) index.
Finally, a global reconstruction of sea level  \citep{Jevrejeva}
and a reconstruction of the North Atlantic Oscillation (NAO) index
\citep{Luterbacher2002} are analyzed and compared: both sequences cover
about three centuries from 1700 to 2000. The proposed methodology quickly
highlights oscillations and teleconnections among the records at the
decadal and multidecadal scales. At the secular time scales tide gauge
records present relatively small (positive or negative) accelerations,
as found in other studies \citep{Houston2011}. On the contrary, from
the decadal to the secular scales (up to 110-year intervals) the tide
gauge accelerations oscillate significantly from positive to negative
values mostly following the PDO, AMO and NAO oscillations. In particular,
the influence of a large quasi 60-70 year natural oscillation is clearly
demonstrated in these records. The multiscale dynamical evolutions
of the rate and of the amplitude of the annual seasonal cycle of the
chosen six tide gauge records are also studied.
\begin{center}
$\sim$
\end{center}
Cite this article as: \textbf{Scafetta, N., 2013.  Multi-scale dynamical analysis (MSDA) of sea level records vs. PDO, AMO, and NAO indexes. \textit{Climate Dynamics}. DOI: 10.1007/s00382-013-1771-3}
\end{abstract}

\section{Introduction}

Understanding the complex dynamics of sea level and tide gauge records
is necessary for testing models of sea level changes against data
and to provide local governments with efficient forecasts for planning
appropriate actions to protect communities from possible sea inundation
and for other civil purposes. Long-term sea level variations are driven
by numerous coupled processes arising from an interaction of eustatic
sea level rise and glacial isostatic subsidence, long-term tidal and
solar cycles, oscillations of ocean circulation, variations in temperature
and/or salinity and other factors that can be also characteristic
of the specific geographical location \citep[e.g: ][]{Douglas1992,Dean2013,Houston2011,Levermann,Morner1989,Morner(1990),Morner,Sallenger,Woodworth1990}.

Despite their intrinsic dynamical complexity, tide gauge records are
often analyzed using simplistic mathematical approaches, such as fitting
one given time interval with a second order polynomial of the type:

\begin{equation}
f(t)=\frac{1}{2}a(t-t_{0})^{2}+v_{0}(t-t_{0})+c_{0},\label{eq:1}
\end{equation}
where $a$ is the sea level average acceleration during the fitted
period (it does not depend on $t_{0}$), $v_{0}$ is the sea level
rate at $t_{0}$, and $c_{0}$ is the average level at $t_{0}$. The
reference date, $t_{0}$, can be changed as necessary.

However, it is easy to demonstrate that fitting tide gauge records within just one generic period
can yield severely ambiguous results. In fact, complex signals are
characterized by specific multi-scale patterns that a regression analysis
on a fixed time interval can obscure. Simply changing the length and
the time period used for the analysis can significantly alter the
values of the regression parameters.

For example, sea-level accelerations are critical components for understanding
the forces that determine their dynamical evolution and for projecting
future sea levels. Without any acceleration, the 20th century
global sea-level average trend of $\sim1.7$ $mm/year$ would produce
a rise of only $\sim153$ $mm$ from 2010 to 2100 \citep{Houston2011}.
However, the scientific literature reports apparently contrasting
results about the acceleration values. Let us discuss a few cases.

(1) \citet{Church2006} used secular-long time series and estimated
that the global sea level from 1880 to 2004 experienced a modest acceleration,
$a=0.013\pm0.006$ $mm/year^{2}$. \citet{Church2011} repeated the
analysis for the period 1880-2009 and found a slightly smaller value,
$a=0.009\pm0.004$ $mm/year^{2}$. \citet{Houston2011} analyzed 25
tide gauges for the period 1930 to 2010 and found a mean negative
acceleration, $a=-0.0123\pm0.0104$ $mm/year^{2}$ (95\%), where the
17 gauge records for the Atlantic zone had an average acceleration
of $a=-0.0138\pm0.0148$ $mm/year^{2}$ (95\%) and 8 gauge records from
the Pacific coast had an average acceleration of $a=-0.0091\pm0.0096$
$mm/year^{2}$(95\%). If the above (positive or negative) acceleration
values persist also for the 21st century, their overall effect would
be modest: the average sea level could rise $\sim150\pm100$ $mm$
from 2010 to 2100 by taking into account also the $\sim1.7$ $mm/year$
average linear rate.

(2) \citet{Sallenger} analyzed the period 1950-2009 and found that
numerous locations of the Atlantic coast of North America are characterized
by strong positive accelerations. Similar results were found also by \citet{Boon}, who used quadratic polynomial regressions to analyze a number of U.S. and Canadian tide gauge records over the 43 years period from 1969 to 2011.  For example, in New York City the
acceleration during the 1950-2009 period was estimated to be $a=0.044\pm0.030$
$mm/year^{2}$ ($1\sigma$ error, annual resolution) \citep[supplementary figure S7]{Sallenger}. For the periods 1960-2009 and 1970-2009 the acceleration values would
progressively increase: $a=0.083\pm0.049$ $mm/year^{2}$ and $a=0.133\pm0.093$
$mm/year^{2}$, respectively. The progressive increase of the acceleration
value was claimed to depart from past values and was interpreted as
due to the anthropogenic warming of the last decades causing significant
changes in the strength of the Atlantic Meridional overturning circulation
and of the Gulf Stream. Evidently, high positive acceleration values
and their progressive increase would be quite alarming if this trend
persists also during the 21st century, as the anthropogenic global
warming theory would predict \citep{IPCC2007}. For example, in New
York City, if the 1970-2009 acceleration remains constant during the
21st century and the 1970-2009 quadratic polynomial fit is used to extrapolate
the 21st century sea level rise, the sea level could increase by $\sim1129\pm480$
$mm$ from 2000 to 2100. For New York City, \citet{Boon} used his 1969-2011 quadratic polynomial fit to project a sea level rise of 390-750 mm above the 1983-2001 sea level mean by 2050. However, I observe that using the 1900-2000 quadratic polynomial fit extrapolation, which produces an acceleration of just $a=0.0032\pm0.0079$
$mm/year^{2}$, in New York City the sea level could increase by just $\sim332\pm60$
$mm$ from 2000 to 2100  (that is mostly due to the linear rate of $v_0=3.16 \pm 0.41$ mm/year), which makes a significant difference.

(3) \citet{Dean2013} analyzed 456 globally distributed monthly tide
gauge records and satellite measured records for the period 1993-2011
and found negative average accelerations (with a relatively large
uncertainty), $a=-0.041$ $mm/year^{2}$ (gauges)
vs. $a=-0.083$ $mm/year^{2}$ (satellites). Therefore,
during the last two decades the sea level accelerations have been
mostly negative at numerous locations. If these  negative
accelerations persist during the 21st century, sea level rates would
significantly slow down and, eventually, sea levels could even decrease
in numerous locations.

(4) \citet{Boretti2012} analyzed two century-long tide gauge records
referring to the east and west coast of Australia, Sydney and Fremantle,
and found secular accelerations of $a=0.014$ $mm/year^{2}$ and $a=-0.0023$
$mm/year^{2}$, respectively, and for the 20-year period from Jan/1990
to Dec/2009 he found accelerations of $a=0.37$ $mm/year^{2}$ and $a=-0.6$8
$mm/year^{2}$, respectively. \citet{Hunter} used the same tide gauge
records used by Boretti but with an annual resolution, and for the
21-year period from 1990 to 2010 found $a=0.44\pm0.34$ $mm/year^{2}$
(Sydney) and $a=-0.60\pm0.70$ $mm/year^{2}$ (Fremantle); they also
reported the global accelerations measured by satellite altimeters
for the 1993-2009 period, $a=-0.027\pm0.114$ $mm/year^{2}$, and from
theoretical computer climate models for the period 1990-2009, $a=0.078\pm0.107$
$mm/year^{2}$.

As it is evident by comparing the above apparently contrasting results,
the evaluated accelerations appear to strongly depend not only on
the location, but also on the time intervals chosen for the regression
analysis. Sea level records are not just randomly evolving around
an acceleration background trending, but they appear to be characterized
by complex natural oscillations. Consequently, each of the above estimated
acceleration values, by alone, may not tell us much about the true
long-range dynamics of tide gauge records. Indeed, those numbers may
be highly ambiguous and may generate confusion to  some readers, e.g. policy
makers, who may wonder whether  sea levels are strongly accelerating,
or strongly decelerating, or not accelerating or decelerating at all.
Moreover, comparing sea level acceleration values from different locations
measured using records of different lengths and  periods \citep[e.g. as done in:][]{Houston2011,Woodworth2009}
may be equally misleading in presence of oscillations.

Indeed, quasi 20-year and 60-year oscillations have been found in
sea level records \citep{Chambers,Jevrejeva,Morner}, although the
patterns referring to individual tide gauge records appear to be quite
complex and inhomogeneous \citep[e.g. ][]{Woodworth2009}. Cyclic
changes could partially explain why the accelerations of these records
are close to zero when quasi secular-long records are analyzed \citep{Church2011,Houston2011},
while the same records may present large positive accelerations when
short periods of 40-60 years between 1950 to 2009 are analyzed, as
done by \citet{Sallenger} and by \citet{Boon}. Cyclic changes could also explain the
large volatility measured in the acceleration values at the decadal
and bi-decadal scales. Indeed, quasi decadal, bi-decadal and 50-90
year oscillations have been found for centuries and millennia in numerous
climate and marine records \citep{Chylek,Klyashtorin,Knudsen,Kobashi,Mazzarella,Morner1989,Morner(1990)}.
Since 1850 quasi decadal, bidecadal and 60-year oscillations have
been found in global temperature records, are typical astronomical/solar
oscillations and will likely exist also in the future \citep{Ogurtsov,Scafetta2010,Scafetta2012m,Scafetta2012b,Scafetta2013,Soon}.
Thus, it is legitimate to expect that equivalent oscillations (with
appropriate phase shifts depending on the geographical location) may
characterize numerous tide gauge records too.

For example, it is possible that \citet{Sallenger} found very large
positive acceleration values in the tide gauge records for New York
City and  other U.S. Atlantic cities (the so-called ``\textit{hot-spot}s'')
simply because these authors compared trends between the periods 1950-1979
and 1980-2009 that could follow from a valley to a peak the 60-year
oscillation commonly found in the Atlantic Multidecadal Oscillation
(AMO), in the Pacific Decadal Oscillation (PDO) and in the North Atlantic
Oscillation (NAO) indexes \citep{Knudsen,Loehle,Mazzarella}. This
60-year oscillation may have made the tide gauge record for New York
City as well as for  other U.S. Atlantic cities concave from 1950 to 2009. A similar critique would apply also to the results by \citet{Boon}, who used 1969-2011 quadratic polynomial fits to project future sea level trends up to 2050.
For the same reason, \citet{Dean2013} found significant negative accelerations
for the period 1993-2011 likely because the bidecadal and the 60-year
oscillations may have been bending down during this period by making
the tide gauge records momentarily convex. Below, I will  demonstrate the validity of the critique.

Herein I propose a multi-scale dynamical analysis (MSDA) methodology
to study dynamical patterns revealed in tide gauge records. MSDA
uses  colored palette diagrams to represent local accelerations,
rates and average annual seasonal amplitudes at multiple scales and
time periods. This methodology aims: (1) to provide a comprehensive
dynamical picture of climatic records; (2) to facilitate their physical
interpretation by highlighting interconnections and dynamical couplings
among climatic indexes; (3) to avoid possible embarrassing ambiguities
emerging from arbitrary choices of the time intervals used for the
regression analysis; (4) to develop more appropriate forecast models
that take into account the real dynamics of these records.

The MSDA visualization technique is used to analyze dynamical details
in six secular-long tide gauge records that approximately provide
a global coverage of the world oceans. Sydney, Fremantle, and New
York City were also analyzed using more simplistic methodologies in
\citet{Boretti2012}, \citet{Hunter} and \citet{Sallenger}: so a
reader may better appreciate the high performance of MSDA by comparing
its results with those recently published by other scientists.  The
MSDA diagrams are compared versus equivalent diagrams referring to
the PDO index and to the AMO index to provide a physical interpretation
of the results. The tide gauge records of Honolulu, San Diego and
Venice are briefly studied to provide a more worldwide picture of
the situation and to highlight additional teleconnection patterns
among different regions. Finally, MSDA diagrams are used to compare
a global reconstruction of sea level \citep{Jevrejeva} and a reconstruction
of the NAO index \citep{Luterbacher}: both sequences cover about
three centuries since 1700 and can be used to highlight a possible physical
coupling and to confirm the presence of a major 60-70 year natural oscillation
in these records.

\begin{table*}
\center
\begin{tabular}{cccc}
\hline
period & Sydney & Fremantle & New York \\
\hline
1/1886 - 12-2010 & $+0.014\pm0.003$ &  &  \\
1/1897 - 12/2010 & $+0.016\pm0.003$ & $-0.007\pm0.006$ & $+0.008\pm0.005$ \\
6/1914 - 12/2010 & $+0.006\pm0.005$ & $-0.018\pm0.009$ & $+0.002\pm0.007$ \\
1/1914 - 12/1993 & $+0.022\pm0.008$ & $-0.080\pm0.014$ & $-0.026\pm0.011$ \\
1/1950 - 12/2009 & $+0.014\pm0.016$ & $+0.083\pm0.029$ & $+0.043\pm0.023$ \\
1/1989 - 12/2010 & $+0.436\pm0.201$ & $-0.054\pm0.360$ & $-0.158\pm0.292$ \\
1/1990 - 12/2009 & $+0.372\pm0.257$ & $-0.678\pm0.458$ & $-0.337\pm0.372$ \\
1/1990 - 12/2010 & $+0.361\pm0.228$ & $-0.598\pm0.400$ & $+0.061\pm0.328$ \\
1/1991 - 12/2010 & $+0.039\pm0.254$ & $-0.739\pm0.459$ & $+0.423\pm0.372$ \\
1/1992 - 12/2010 & $-0.233\pm0.287$ & $-0.920\pm0.520$ & $+0.676\pm0.435$ \\
1/1993 - 12/2010 & $-0.549\pm0.327$ & $-0.920\pm0.600$ & $+0.923\pm0.497$ \\
\hline
\end{tabular}
\caption{Tide gauge level accelerations ($mm/year^{2}$) in Sydney, Fremantle
and New York within arbitrary time intervals using Eq. \ref{eq:1}.}
\end{table*}

\begin{table*}
\center
\begin{tabular}{cccc}
\hline
period & Eq. 1 - year & Eq. 1 - month & Eq. 2 - month \\
\hline
1/1893 - 12/2010 & $0.0102\pm0.0052$ & $+0.0078\pm0.0042$ & $+0.0073\pm0.0031$ \\
1/1900 - 12/2010 & $0.0075\pm0.0062$ & $+0.0046\pm0.0049$ & $+0.0041\pm0.0037$ \\
1/1910 - 12/2010 & $0.0028\pm0.0078$ & $-0.0008\pm0.0061$ & $-0.0014\pm0.0046$ \\
1/1920 - 12/2010 & $0.0005\pm0.0102$ & $-0.0042\pm0.0079$ & $-0.0048\pm0.0060$ \\
1/1930 - 12/2010 & $0.0053\pm0.0138$ & $-0.0013\pm0.0106$ & $-0.0020\pm0.0081$ \\
1/1940 - 12/2010 & $0.0414\pm0.0188$ & $+0.0334\pm0.0148$ & $+0.0328\pm0.0114$ \\
1/1950 - 12/2010 & $0.0687\pm0.0286$ & $+0.0583\pm0.0216$ & $+0.0583\pm0.0169$ \\
1/1960 - 12/2010 & $0.1203\pm0.0455$ & $+0.1069\pm0.0341$ & $+0.1086\pm0.0265$ \\
1/1970 - 12/2010 & $0.1941\pm0.0830$ & $+0.1725\pm0.0601$ & $+0.1797\pm0.0469$ \\
1/1980 - 12/2010 & $0.0644\pm0.1762$ & $+0.0163\pm0.1222$ & $+0.0413\pm0.0950$ \\
1/1990 - 12/2010 & $0.2223\pm0.4678$ & $+0.0606\pm0.3284$ & $+0.1096\pm0.2553$ \\
1/2000 - 12/2010 & $2.2596\pm1.643$ &  $+2.5584\pm1.761$ &  $+2.1697\pm1.249$ \\
\hline
\end{tabular}
\caption{Analysis of the acceleration error-bars for the tide gauge record
of New York City within different intervals and using: (1) Eq. \ref{eq:1}
and the annual average record; (2) Eq. \ref{eq:1} and the monthly
average record; (3) Eq. \ref{eq:1-1} and the monthly average record.
The methodology (3) produces the smallest error-bars. }
\end{table*}

\section{Data}

Tide gauge records were downloaded from the Permanent Service for
Mean Sea Level (PSMSL) \citep{Woodworth(2003),PSMSL2013}. The following
six monthly resolution records are used: Sydney (Fort Denison, ID
\#65, Jan/1886 - Dec/1993; Fort Denison 2, ID \#196, Jun/1914 - Dec/2010);
Fremantle (ID \#111, Jan/1897 - Dec/2010); and New York City (the
Battery, ID \#12, Jan/1856 - Dec/2011); Honolulu (ID \#155, Jan/1905
- Dec/2011); San Diego (ID \#158, Jan/1906 - Dec/2011); and Venice
(ID \#168, Jan/1909 - Dec/2000). The two records from Sydney are practically
identical during the overlapping period and are combined to form a
sequence from 1886 to 2010.

The following monthly records are also used: the Pacific Decadal Oscillation
(PDO) index (Jan/1900 - Jan/2013);
the Atlantic Multidecadal Oscillation (AMO) index (Jan/1856 - Jan/2013).
The PDO and AMO indexes represent modes of variability occurring in
the Pacific Ocean and in the North Atlantic Ocean, respectively, which
have their principal expression in the sea surface temperature field
once linear trends are removed. The oscillations revealed in these
records likely arise from a quasi-predictable variability of the ocean\textendash{}atmosphere
system observed for centuries and millenia \citep{Klyashtorin,Knudsen,Mantua,Scafetta2010,Scafetta2012b,Schlesinger},
which is expected to influence tide gauge records as well given their dependency on global temperature related climatic changes.

Finally, a global reconstruction of sea level  \citep{Jevrejeva}
and a reconstruction of the North Atlantic Oscillation (NAO) \citep{Luterbacher,Luterbacher2002}
are also studied: both sequences cover about three centuries from 1700 to 2000. The NAO index is defined as the average air pressure
difference between a low latitude region (e.g. Azores) and a high
latitude region (e.g. Iceland). These longer records are used to demonstrate
the existence of a quasi 60-70 year natural oscillation, which needs
to be taken into account for correctly interpreting the results obtained by
using multidecadal-long tide gauge records.

Let us run a test to demonstrate that fitting tide gauge records within fixed
periods, as done in the studies summarized in the Introduction, yields
severely ambiguous results.  I used Eq. \ref{eq:1} to
fit generic intervals and the results referring to Sydney, Fremantle
and New York City are reported in Table 1. The acceleration values
change greatly from significantly positive to significantly negative
values. In particular, at the bidecadal scales the variation of the
acceleration can be quite significant and sudden. For example, for
Sydney a strong positive acceleration, $a=+0.36\pm0.23$ $mm/year^{2}$,
is found for the 21-year period from Jan/1990 to Dec/2010 \citep[as found also in][]{Hunter},
but a strong negative acceleration, $a=-0.55\pm0.32$ $mm/year^{2}$,
is found for the 18-year period from Jan/1993 to Dec/2010. The results
obtained for Fremantle are also interesting: always negative acceleration
values are found with the exception of the 60-year period 1950-2009 \citep[which is the period chosen by][]{Sallenger}
when the acceleration is significantly positive. Thus, tide gauge
records are statistically and dynamically complex, and estimating
a single acceleration value within an arbitrarily chosen time interval,
in particular at scales shorter than 60-70 years, can give extremely
ambiguous and misleading results.

\section{Multi-scale dynamical analysis (MSDA) of tide gauge records}

To reduce the ambiguities highlighted above, tide gauge rates and
accelerations have to be computed using moving time windows of different
length. Recently \citet{Parker2012a,Parker2012b,Parker} used moving
time windows of 20, 30 and 60 years to demonstrate the presence of
oscillations in tide gauge records at the chosen scales. However,
a comprehensive dynamical picture can be provided by analyzing all
available scales at once. For example, \citet[figure 2]{Jevrejeva}
calculated acceleration coefficients using variable windows (from 10
to 290 years) of their global sea level record. This multi-scale technique
and its graphical representation will be herein extended, applied
to local and global sea level records and used for comparisons against
the PDO, AMO and NAO indexes. The proposed methodology aims to provide
sufficiently compact and eye-catching diagrams to facilitate the study
of these records and to rapidly evaluate, even visually, dynamical
patterns and cross-correlations among alternative climatic
records that can be physically coupled and/or teleconnected to each
other.

Records with a monthly resolution are adopted because the statistical
error for the regression coefficients is normally smaller than using
the annual resolution records due to the fact that the monthly data
are not Gaussian-distributed around an average trend but are autocorrelated
because constrained by the annual seasonal cycle \citep[e.g.: ][figure 3]{Church2011}.
However, the mean regression coefficients do not change significantly
using the annual instead of the monthly resolution, in particular
for longer scales. In any case, because the monthly tide gauge records
are characterized by a strong annual seasonal cycle, Eq. \ref{eq:1}
should be substituted with

\begin{figure}[!]
\includegraphics[width=1\columnwidth,height=0.5\textheight]{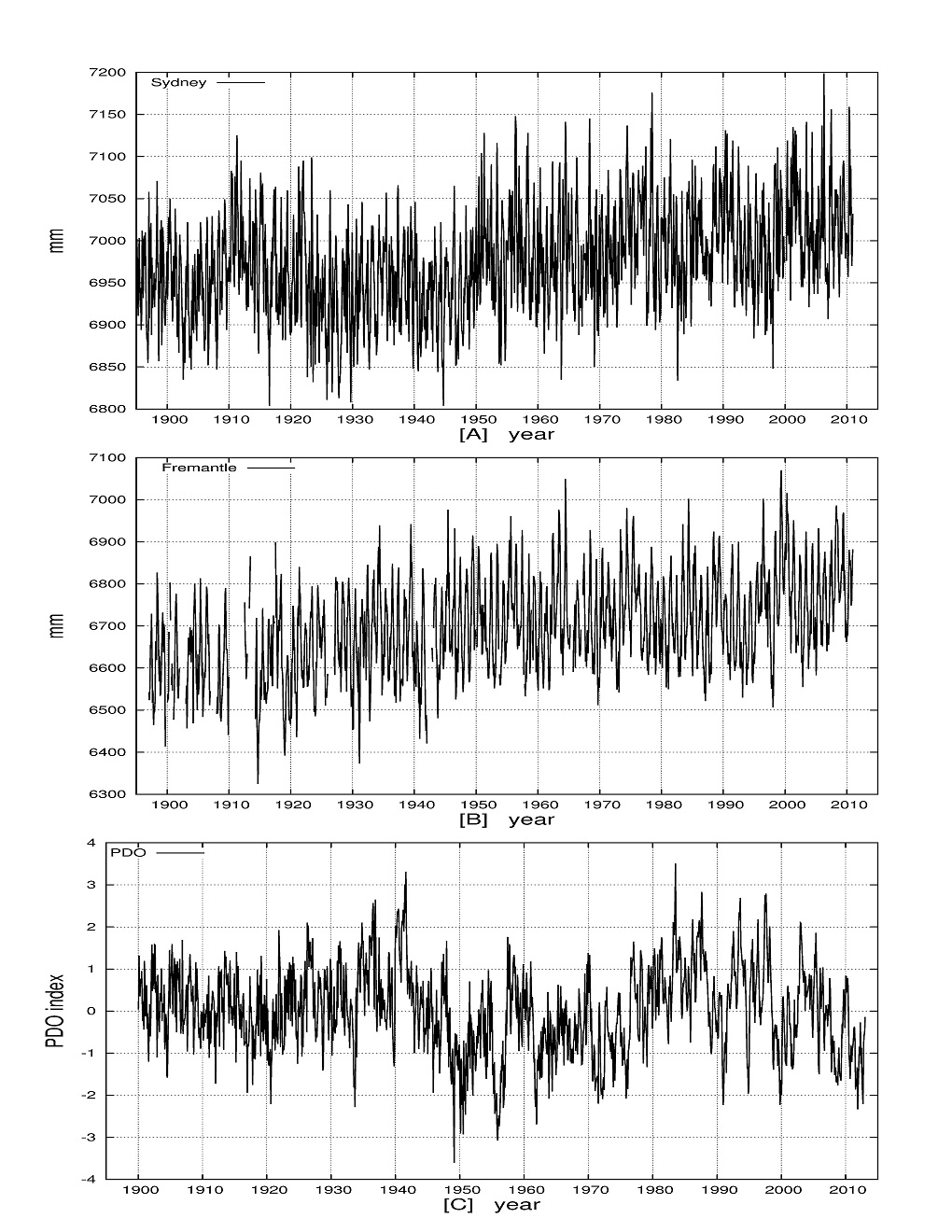}\caption{[A] Tide gauge record in Sydney. [B] Tide gauge record in Fremantle. [C] The Pacific Decadal Oscillation (PDO) index. }
\includegraphics[width=1\columnwidth]{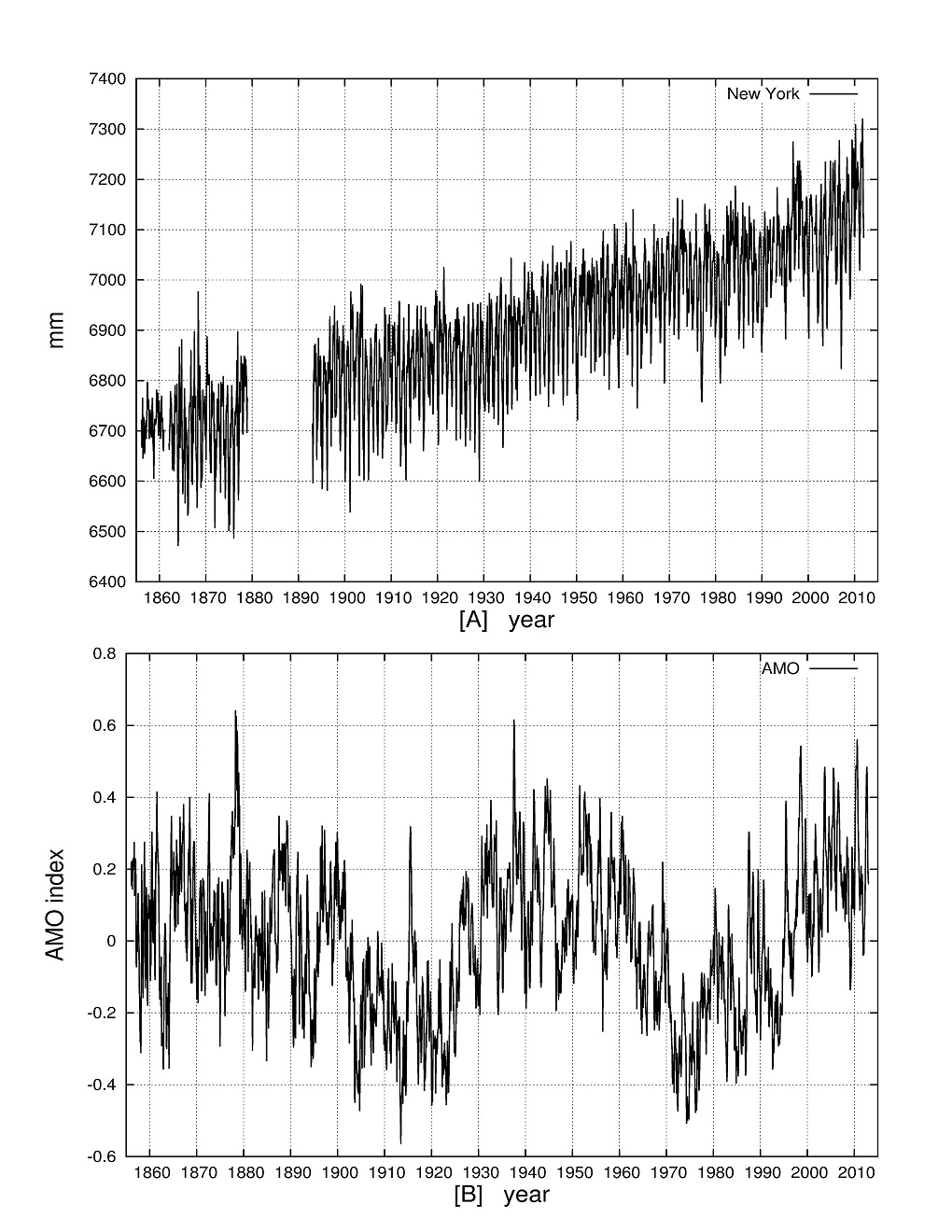}\caption{[A] The tide gauge record in New York. [B] The Atlantic Multidecadal Oscillation (AMO) index. }
\end{figure}

\begin{figure}[!]
\includegraphics[width=1\columnwidth,height=0.5\textheight]{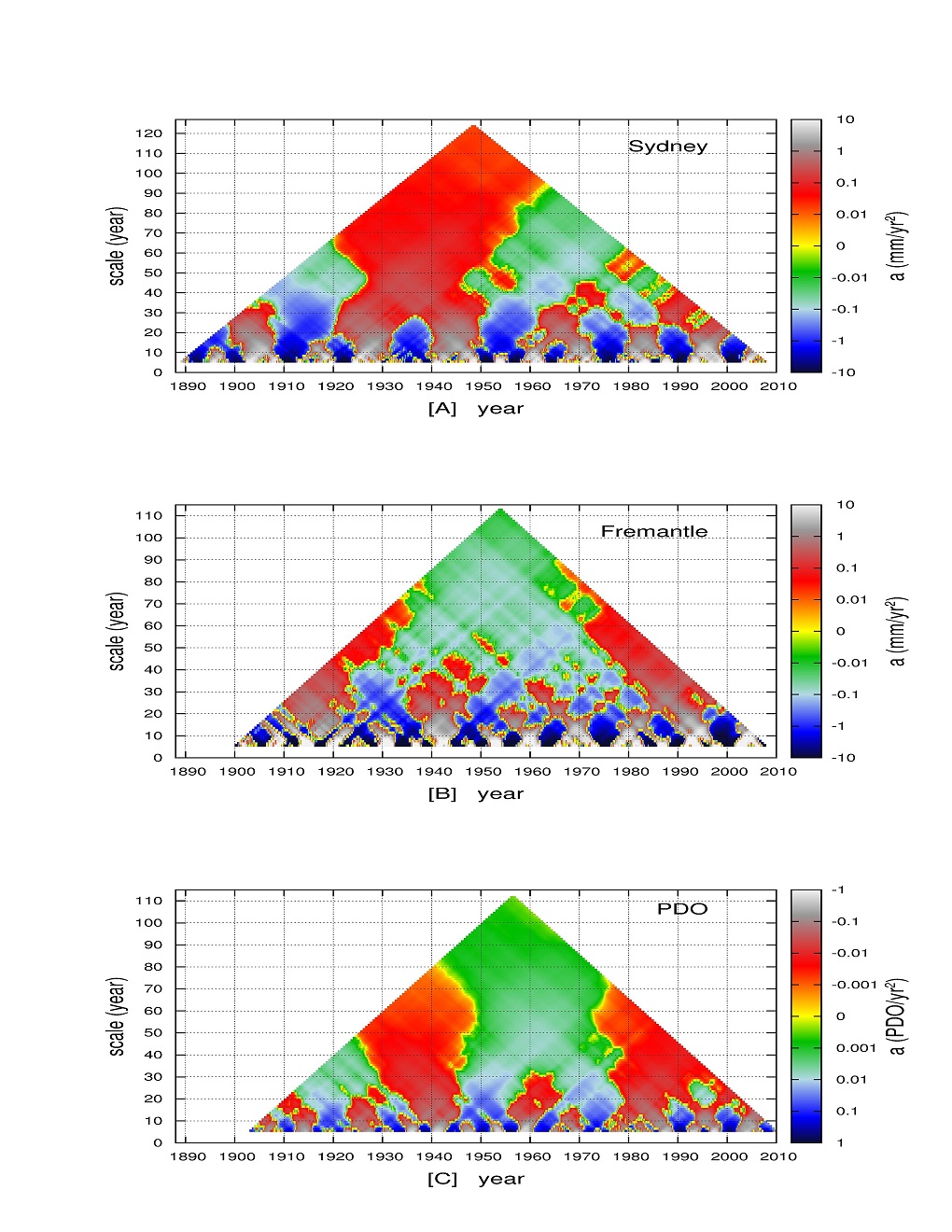}\caption{MSAA colored diagrams for: [A] the tide gauge record in Sydney; [B] the tide gauge record in Fremantle; [C] for the Pacific
Decadal Oscillation (PDO) index (here the colors are inverted). }
\includegraphics[width=1\columnwidth]{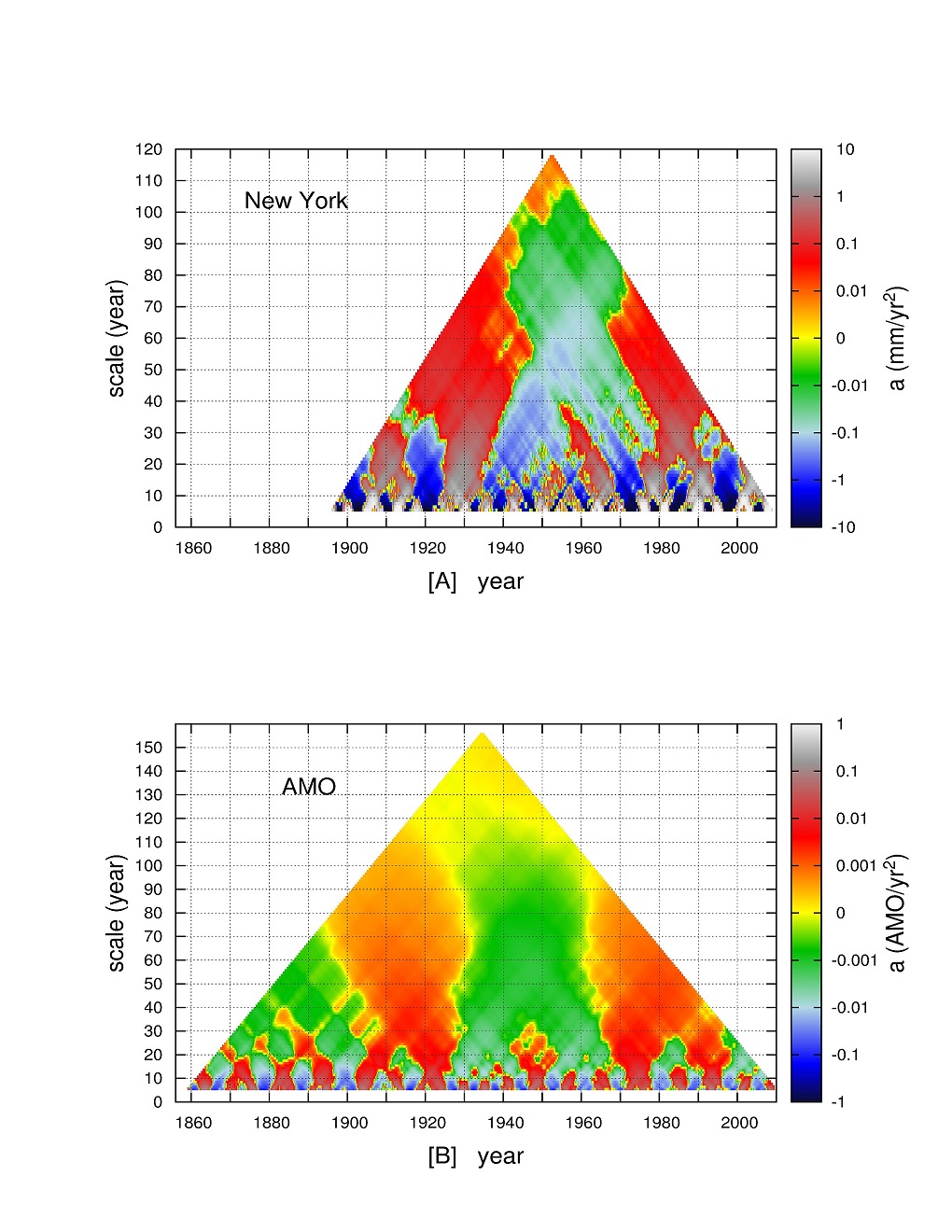}\caption{MSAA colored diagrams for: [A] the tide gauge record in New York; [B] the Atlantic Multidecadal Oscillation (AMO) index. }
\end{figure}

\begin{equation}
f(t)=\frac{1}{2}a(t-t_{0})^{2}+v_{0}(t-t_{0})+c_{0}+H_{c}\cos(2\pi t)+H_{s}\sin(2\pi t),\label{eq:1-1}
\end{equation}
which takes into account the annual seasonal cycle too and could further
reduce the regression uncertainties.

A test was run to evaluate how the regression error associated to
the acceleration depends on different regression methodologies. Table
2 shows the difference of using: (1) the annual resolved records fit
with Eq. \ref{eq:1} \citep[e.g. as used in ][]{Hunter}; (2) the
monthly resolved records fit with Eq. \ref{eq:1}; (3) the monthly
resolved records fit with Eq. \ref{eq:1-1}. The acceleration results
reported in Table 2 indicate that the methodology (1) provides the
worst results (= the largest error bars) at all time scales; the methodologies
(2) and (3) produce almost identical average values although in the
case (3) the regression algorithm would produce smaller error bars.
Adopting the methodology (3) instead of the methodology (1) reduces
the error bars by $\sim40\%$. Thus, the methodology (3) should be
preferred for tide gauge records if  high precision is required for
specific purposes.

\subsection{Multi-scale acceleration analysis (MSAA)}

The multi-scale acceleration analysis (MSAA) methodology evaluates
the acceleration values using Eq. \ref{eq:1-1} for all possible available
time window lengths, which are called ``scales'', and plot the acceleration
results in a colored palette diagram in function of the time position
of the regression interval (abscissa) and of the scale (ordinate).
The pictures are then studied for dynamical patterns and compared
to equivalent diagrams referring to other records to identify possible
dynamical correlations, couplings, coherences and teleconnections.
No significant visible changes would be obtained by adopting Eq. \ref{eq:1}.

Figure 1 shows the tide gauge records for Sydney and Fremantle vs.
the PDO index. Figure 2 shows the tide gauge records for New York
City vs. the AMO index. Figure 3 depicts the MSAA diagrams for the
tide gauge records of Sydney and Fremantle, and compare them with
that for the PDO index. In the latter case the colors are inverted
for reasons explained below. Figure 4 depicts the MSAA diagrams for
the tide gauge record of New York City (since 1893) and for the AMO
index. Time scales from a minimum of 5-year periods to the total length
of the data are used. The 5-year period is just a lower limit, which
is chosen because it allows to detect the decadal oscillations that
are known to exist in the climate system \citep[e.g.: ][]{scafetta2009,Scafetta2010,Scafetta2012m,Manzi}.

\begin{figure*}[!t]
\includegraphics[width=1\textwidth]{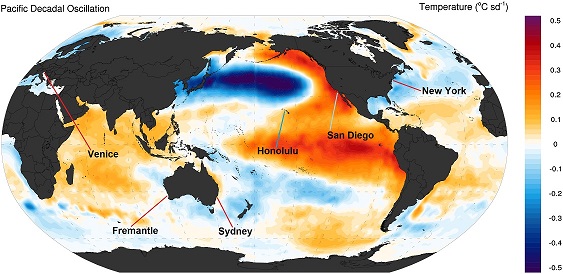}\caption{Pacific Decadal Oscillation (PDO), positive phase and global pattern. (Picture adapted from wikipedia, \protect\href{http://en.wikipedia.org/wiki/Pacific_decadal_oscillation}{http://en.wikipedia.org/wiki/Pacific\_{}decadal\_{}oscillation}).
The figure shows also the locations of the six chosen sites studied in the paper.}
\end{figure*}

The MSAA diagrams for the sea level records use a logarithmic scale
for accelerations $|a|\geq0.001$ $mm/year^{2}$ and a linear scale
for $|a|\leq0.001$ $mm/year^{2}$. The MSAA diagrams for the PDO and
AMO indexes use a logarithmic scale for accelerations $|a|\geq0.0001$
$mm/year^{2}$ and a linear scale for $|a|\leq0.0001$ $mm/year^{2}$.
Alternative scales for the acceleration, e.g. a cube root scale, can
be adopted but do not give better results. The ordinate axis for the
time interval scale adopts a linear scale, which produces triangle-shaped
MSAA diagrams; however, a log-scale can be adopted for longer sequences
and/or for highlighting the shortest scales as done in Section 4.

The interpretation of the diagrams is as follows. The color at specific
coordinates, let us say at $x=1940$ and $y=40$, gives the acceleration
calculated using Eq. \ref{eq:1-1} to fit the data for the 40-year
long interval (scale $y=40$ years) from 1920 to 1960 (period centered
in $x=1940$), and so on. Orange to red to white colors indicate incrementally
positive accelerations, while green to blue to black colors indicate
incrementally negative accelerations. Yellow colors indicate acceleration
values very close to zero. The accelerations for the tide gauges are
calculated in $mm/year^{2}$, while for the PDO and AMO indexes the
units are those of these indexes, which are temperature related, per
$year^{2}$.

The statistical error bars associated to the single acceleration measurements
depend on the specific tide gauge record and on the scale. As Table
2 show, the error bars may vary from about $\pm0.003$ $mm/year^{2}$
at the secular scale to about $\pm1$ $mm/year^{2}$ at the decadal
one using Eq. \ref{eq:1-1} with monthly records. However, when multiple
acceleration values corresponding to a single scale are compared to
determine a dynamical pattern, the average statistical error associated
to it becomes smaller than the statistical error associated to each
single acceleration value by a factor equal to the root of the number
of degrees of freedom that may be approximately estimated to be equal
to the ratio between the length of the data and the analyzed scale:
e.g. for a 100-year sequence the error in the dynamical pattern may
vary from about $\pm0.003$ $mm/year^{2}$ at the secular scale to about $\pm0.3$
$mm/year^{2}$ at the decadal one. This error bars are relatively small
relatively to the maximum and minimum acceleration values observed
at each scale. Therefore, the observed patterns cannot emerge from
just statistical noise but are dynamically relevant. Moreover, as
demonstrated below, similar patterns emerge in alternative records,
which further demonstrates their physical origin.

The diagrams for the tide gauge records indicate: (1) these records
are characterized by alternating periods of positive and negative
accelerations at the decadal and multidecadal scales; (2) at the decadal
and bi-decadal scales, strong quasi periodic oscillations are observed,
as indicated by the quasi-periodic shifts from deep green regions
to deep red regions; (3) at larger scales from about  30-year to 110-year,
the accelerations are moderately positive both at the beginning and
at the end of the records (in particular from Fremantle and New York),
while in the middle of the records large regions with negative accelerations
are observed, this pattern may indicate the presence of a large multidecadal
oscillation that will be better discussed in Section 4; (4) the accelerations converge to small values close
to zero (green-yellow-orange color) at the secular scale, which is
indicated at the top of the triangle.

\begin{figure}[!t]
\includegraphics[width=1\columnwidth]{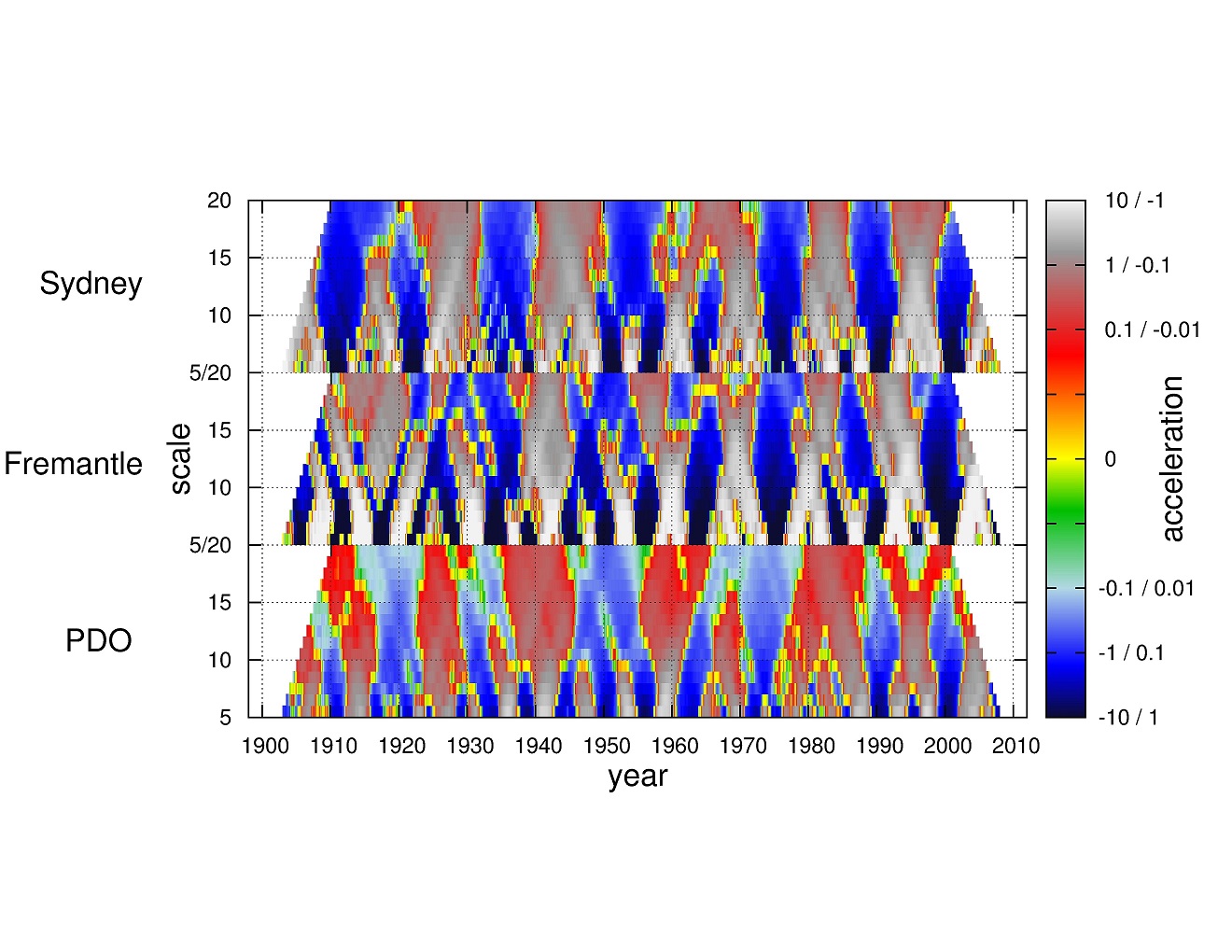}\caption{Zoom of Figure 3 within the scales from 5 to 20 years to highlight the good cross correlation among the three records at the decadal/bidecadal
scales. Table 3 reports the cross-correlation coefficients referring
to each annual scale. }
\end{figure}

\begin{figure*}[!t]
\includegraphics[width=1\textwidth,height=0.7\textheight]{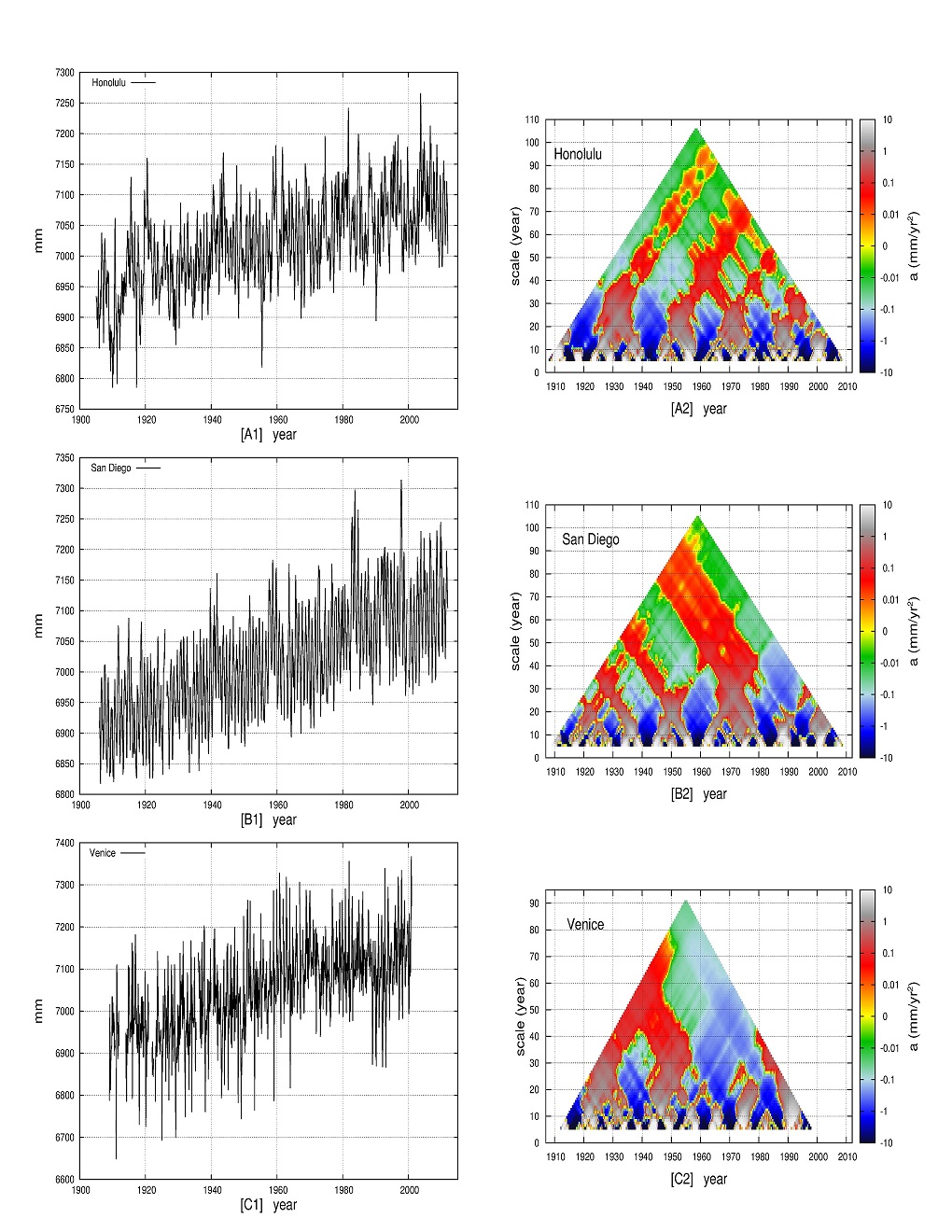}\caption{Tide gauge records (left) and MSAA colored diagrams (right) for: [A] Honolulu, U.S. state of Hawaii; [B] San Diego, U.S. state of California;
 [C] Venice, Italy. }
\end{figure*}

The diagram referring to the PDO index (Fig. 3C) is plotted with inverted
colors (that is, green/blue is used for positive accelerations while
orange/red for negative accelerations) because when the PDO is in its ``warm'',
or ``positive'' phase, the west South Pacific (e.g. around Australia)
cools while part of the eastern Pacific ocean warms; the opposite
pattern occurs during the ``cool'' or ``negative'' PDO phase.
For example, Figure 5 shows the PDO positive phase global pattern.
So, during the PDO warm phase, the sea level around Australia may
decrease, and the accelerations patterns between PDO and tide gauges
in Australia may be negatively correlated.

\begin{figure*}[!t]
\includegraphics[width=1\textwidth,height=0.7\textheight]{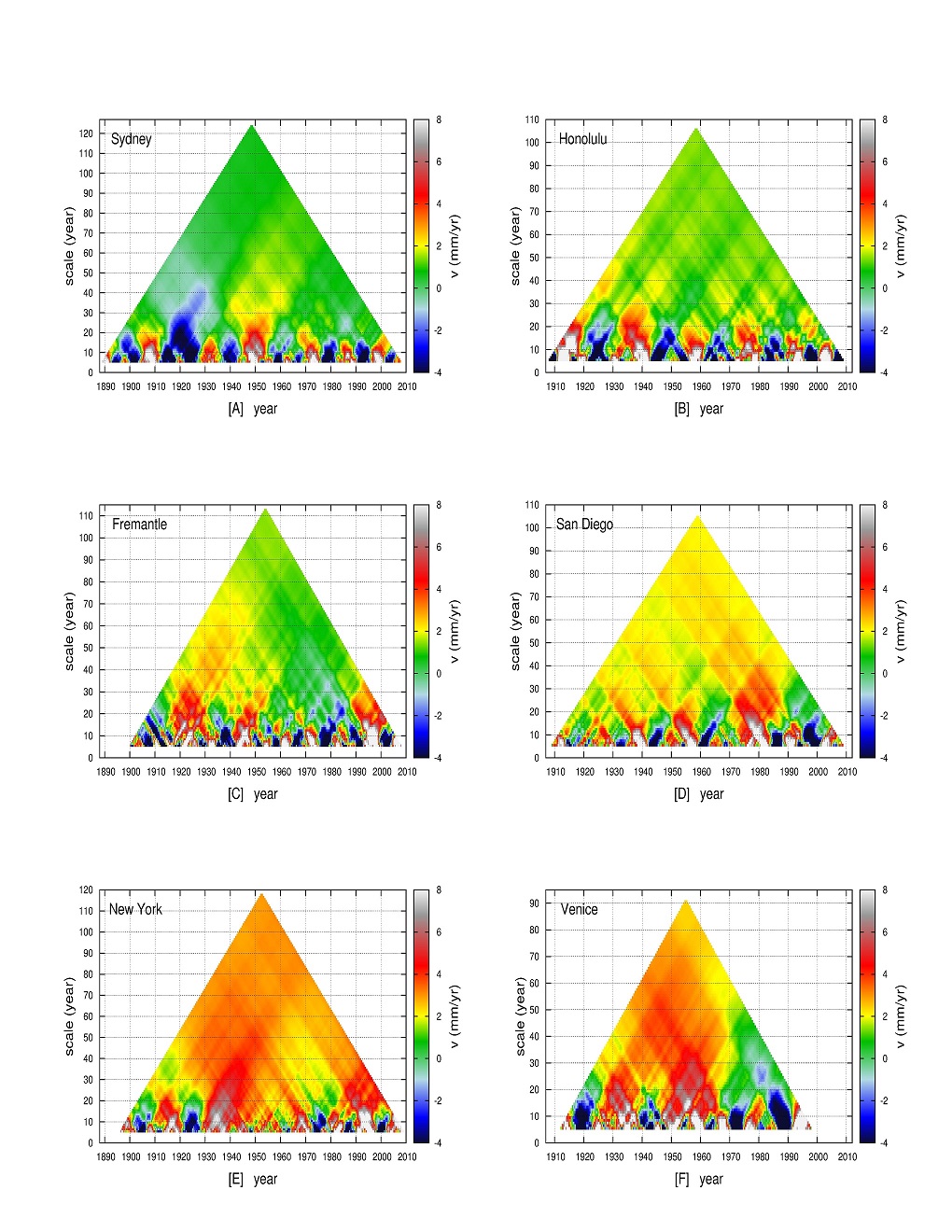}\caption{MSRA colored diagrams for: [A] Sydney, Australia (East coast); [B] Honolulu, U.S. state of Hawaii; [C] Fremantle, Australia (West coast); [D] San Diego, U.S. state of California; [E] New York, U.S. state of New York; [F] Venice, Italy. }
\end{figure*}

The comparison between Sydney (Fig. 3A) and PDO (Fig. 3C) suggests a good
negative (this is because the PDO diagram uses inverted colors) correlation,
which is indicated by a same blue and red region correspondence at
the 10-30 year scales. For example, note the same blue areas in both
the sea level diagrams and the PDO diagram around 1910, 1920, 1930-1940,
1950-1960, 1970-1980, 1990 and 2000. At larger scales the correlation
pattern is more uncertain and, as Fig. 3A shows, for Sydney the correlation
with the PDO patterns becomes positive at higher scales. However,
in general, long-range patterns in tide gauge records may be driven
by numerous physical processes, not just the temperature related ones,
but studying the precise cause of an observed pattern is beyond the
purpose of the present work.

\begin{figure*}[!t]
\includegraphics[width=1\textwidth,height=0.7\textheight]{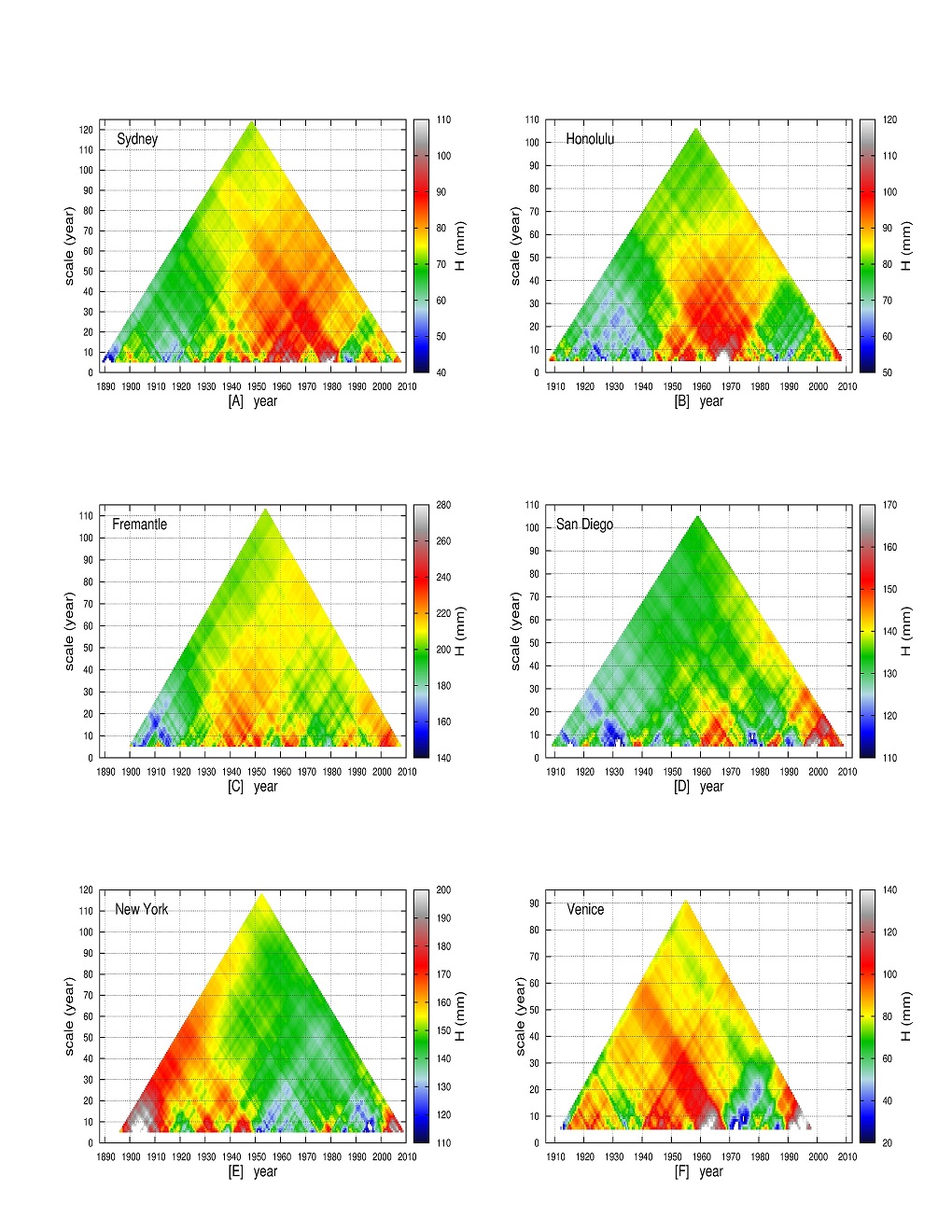}\caption{MSACAA colored diagrams for: [A] Sydney, Australia (East coast); [B] Honolulu, U.S. state of Hawaii; [C] Fremantle, Australia (West coast); [D] San Diego, U.S. state of California; [E] New York, U.S. state of New York; [F] Venice, Italy. }
\end{figure*}

The comparison between Fremantle (Fig. 3B) and PDO (Fig. 3C) suggests
extended (negative) correlations (indicated by a same color area correspondence)
at almost all scales, in particular since 1930. The influence of the
quasi 60-year oscillation is evident. There appears to be a small
phase shift that depends on the scale. In addition, since 1930 the PDO
index may be more accurate because before 1930 the uncertainties in sea
surface temperature records are significantly higher and, in general,
the uncertainties are smaller for the Indian and Atlantic Oceans than
for the Pacific Ocean \citep{Kennedy,Kennedyb}.

\begin{table}[!t]
\center
\begin{tabular}{ccc}
\hline
scale (year) & PDO-Fremantle & PDO-Sydney \\
\hline
5 & -0.44  & -0.18  \\
6 & -0.43  & -0.25  \\
7 & -0.45  & -0.35  \\
8 & -0.44  & -0.41  \\
9 & -0.38  & -0.37  \\
10 & -0.29  & -0.42  \\
11 & -0.31  & -0.37  \\
12 & -0.36  & -0.40  \\
13 & -0.38  & -0.41  \\
14 & -0.38  & -0.38  \\
15 & -0.35  & -0.30  \\
16 & -0.27  & -0.23  \\
17 & -0.24  & -0.22  \\
18 & -0.26  & -0.22  \\
19 & -0.30  & -0.24  \\
20 & -0.32  & -0.20  \\
\hline
\end{tabular}
\caption{Cross-correlation coefficient $r$ between annual MSAA lines for:
(1) PDO and Fremantle; (2) PDO and Sydney. See Figure 6.}
\end{table}

Figure 6 zooms and compares the three MSAA diagrams for Sydney, Fremantle
and PDO within the scales from 5 to 20 years. Extended cross-correlations
of the two tide gauge records with the PDO index are evident. The
cross-correlation coefficient between Sydney and PDO MSAA diagram
areas within the chosen 5-20 year scale is $r=-0.31$ and between
Fremantle and PDO is $r=-0.37$. These cross-correlation values are
highly significant considering the length of the records. See Table
3 for cross-correlation values for each annual scale.

The comparison between New York (Fig. 4A) and AMO (Fig. 4B) MSAA diagrams
suggests extended correlations at almost all scales, although an evident
time-lag of about 10-15 years is observed at the scales from 30
to 110 years. This scale range also highlights the strong influence
of the 60-year AMO oscillation on this tide gauge record. Time-lags
between the AMO and PDO indexes and tide gauge records depend on the
specific location because a change of the temperature may induce variation
in the ocean circulation that effects differently each location \citep[e.g.:][]{Woodworth2009}.

Figure 7 shows the data and the MSAA diagrams for the tide gauge records
of Honolulu, San Diego and Venice. All three records present a negative
acceleration at the secular scale (green color at the top of the triangles).
At the 10-30 year scales, Honolulu and San Diego present large and
synchronous quasi bi-decadal oscillations (the color alternate between
blue and red every about 20 years). At scales larger than 30 years
the MSAA for Honolulu and San Diego reveals interesting symmetric
patterns with green and red oblique bands moving in opposite directions.
This pattern may emerge because according to Figure 5 during the PDO
warm phase San Diego warms while Hawaii is located in a boundary region
of the Pacific Ocean between the warm and the cold regions. The tide
gauge record for Venice presents some similarity with the other records
at the 10-30 year scales with clear decadal and bi-decadal oscillations,
in particular with the AMO index (Fig. 4B). In Venice the sea level
acceleration was moderately positive during the beginning of the 20th
century, while it became moderately negative at the end of the 20th
century for scales larger than 30 years.

An extended quantitative analysis of the time-lags and cross-correlation
patterns among different records, which may also be scale dependent,
is left to future research and it is not relevant for this work. A
visual cross-correlation study among different records and scales
can be in first approximation easily accomplished on a computer screen
by drawing a vertical line crossing the MSAA diagrams (which in the
used figures are carefully alined) and moving the line horizontally
to analyze the areas of interest.

\begin{figure*}[!t]
\includegraphics[width=1\textwidth,height=0.5\textheight]{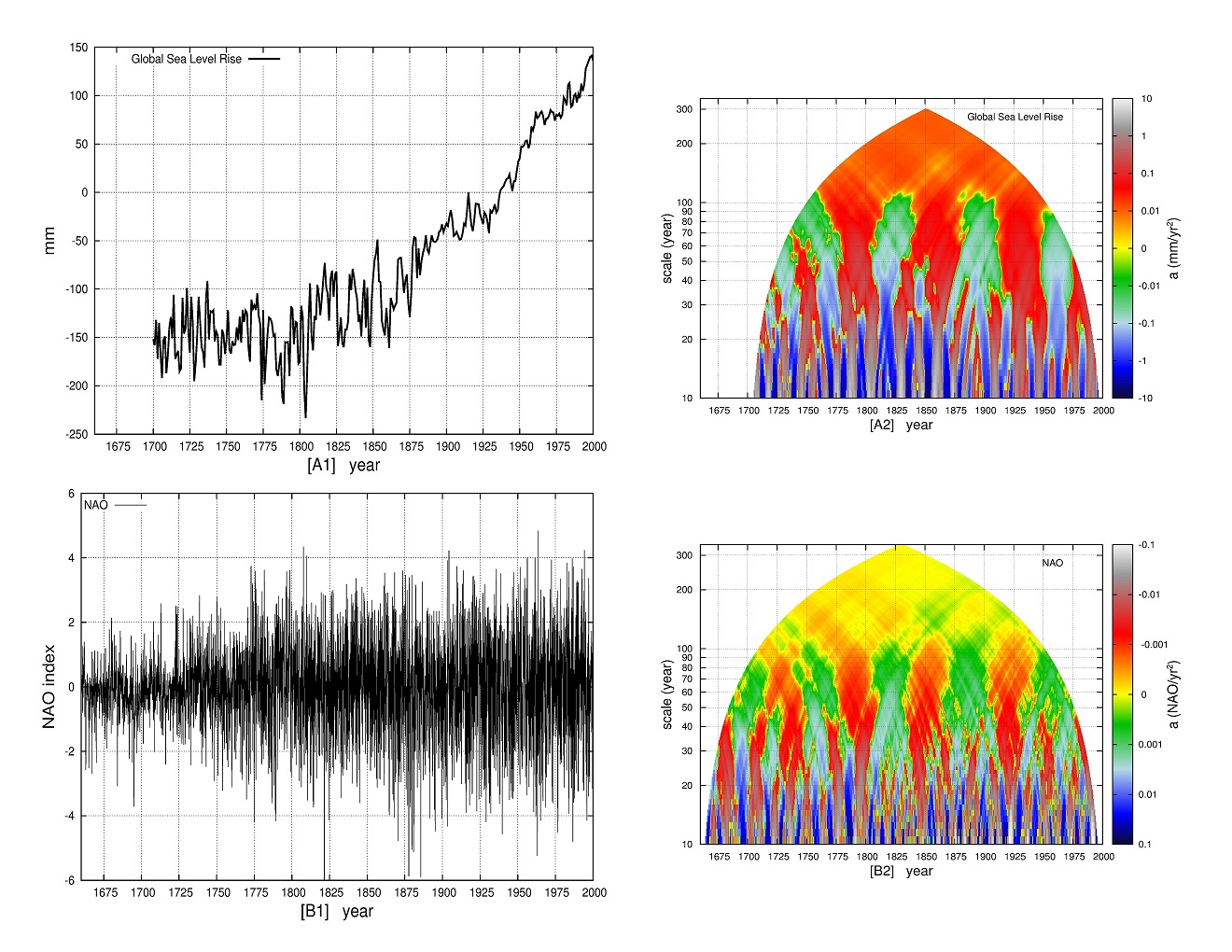}\caption{[A] Global sea level record \citep{Jevrejeva} (left) and its MSAA colored diagram (right). [B] North Atlantic Oscillation (NAO)
\citep{Luterbacher,Luterbacher2002} (left) and its MSAA colored diagram (right). In [B] the colors are inverted. Note the common 60-70 year oscillation since 1700 indicated by the alternating green and red regions within
the 30-110 year scales. }
\end{figure*}

\subsection{Multi-scale rate analysis (MSRA)}

The ``acceleration'' used in the MSAA can be substituted with other
physical measures (e.g., rates, etc.) as necessary. Here a multi-scale
rate analysis (MSRA) is proposed. For example, for determining the
average rate, $v$, during a specific time period, the data have to
be fit with a linear function of the type:

\begin{equation}
g(t)=v(t-t_{0})+c_{0}+H_{c}\cos(2\pi t)+H_{s}\sin(2\pi t),\label{eq:2}
\end{equation}
where $v$ is the average rate in $mm/year$ during the analyzed period,
and $c_{0}$is the level at $t_{0}$. The annual oscillation is added
in the regression model for improved accuracy although almost no difference
in the colored diagrams would be observed for tide gauge records.
Once the rate values are calculated for all available time intervals
and scales, they can be plotted using a colored palette diagram. Figure
8 shows the results obtained for the six tide gauge records used in
the paper.

Evidently, also the rate changes and oscillates in function of the
time interval and of the scale. At smaller scales (10-30 year)  a
large volatility is observed, and also negative rates are observed
during specific periods. Extended periods with rates close to zero
(green color) are observed in many records.

The color at the top of the triangles gives the average secular rate
at the highest available scale, that is, for the entire available
record. The following secular rate values are obtained: {[}A{]} Sydney,
$v=0.63\pm0.04$ $mm/year$; {[}B{]} Honolulu, $v=1.44\pm0.04$ $mm/year$;
{[}C{]} Fremantle, $v=1.51\pm0.09$ $mm/year$; {[}D{]} San Diego, $v=2.04\pm0.06$
$mm/year$; {[}E{]} New York, $v=2.99\pm0.06$ $mm/year$; {[}F{]} Venice,
$v=2.4\pm0.1$ $mm/year$.

\subsection{Multi-scale annual cycle amplitude analysis (MSACAA)}

Eq. \ref{eq:1-1} contains harmonic constructors that can be used
to evaluate the average amplitude and phase of the annual cycle presents in
the tide gauge records. For example, for each scale and period it is possible to
evaluate a total average amplitude (from minimum to maximum) with
the equation:

\begin{equation}
H=2\sqrt{H_{c}^{2}+H_{s}^{2}}.
\end{equation}
The MSACAA diagrams for the six tide gauge records are shown in Figure
9.

Also the average seasonal annual amplitude of the tide gauge records
varies with the scale and the time period. In the case of Sydney,
Hononulu, Fremantle and Sydney, which are within the area of influence
of the Pacific Ocean, the amplitude of the annual cycle increased
on average almost at all scales above 20 years from 1900 to 2000,
as demonstrated by the predominantly green colors on the left side
of the diagrams that slightly become yellow/red on the right side.
However, the opposite pattern is observed for New York City and Venice,
which are within the area of influence of the Atlantic Ocean. Thus,
the analysis appears to reveal an important asymmetric and compensating
dynamical evolution between the Pacific and the Atlantic oceans. Extending
the analysis to additional records may clarify this issue. At smaller
scales, several oscillations are observed.

\section{Multisecular MSAA comparison between the global sea level record
and the NAO index}

Figure 10A depicts a global sea level record \citep{Jevrejeva} (left)
and its MSAA (right). Eq. \ref{eq:1} is used because the record has
an annual resolution. Figure 10B depicts a reconstruction of the North
Atlantic Oscillation (NAO) \citep{Luterbacher,Luterbacher2002} (left)
and its MSAA (right). Both records cover about 300 years from 1700
to 2002 and from 1658 to 2001, respectively. In Figure 10B, which
refers to the NAO index, the colors of the MSAA diagram are inverted.
Note that \citet[figure 2]{Jevrejeva} used a linear scale truncated
at $\pm0.03$ $mm/year^{2}$, while I propose to use a logarithmic scale
to cover a wider and more comprehensive range of $\pm10$ $mm/year^{2}$.

It is easy to notice that the two records present well correlated
quasi 60-70 year oscillations since 1700, as indicated by the alternating
green and red regions: there appear to be a small phase-lag of a few
years. This result indicates that at the 30-100 year time scales the
local accelerations of this global sea level record and of the NAO
index are negatively correlated. Thus, in the global sea level record
proposed by \citet{Jevrejeva} the sea level decreases when NAO is
positive, that is the air pressure at the lower latitudes is high.

The following cross-correlation coefficients between the two MSAA
area diagrams are found: within the 20-100 year scales, $r=-0.29$;
within the 30-100 year scales, $r=-0.44$; within the 40-100 year
scales, $r=-0.57$. These correlation values are highly significant.

For scales larger than 110 years up to 300 years, the sea level presents
quasi uniform accelerations of about $0.01$ $mm/year^{2}$ from 1700
to 2000, as indicated by the orange color in Figure 10A2. For example,
fitting the global sea level record during the 1700-2000 period gives
$a=0.0092\pm0.0004$ $mm/year^{2}$, during the preindustrial 1700-1900
period gives $a=0.0093\pm0.0013$ $mm/year^{2}$, and during the industrial
1900-2000 period gives $a=0.010\pm0.004$ $mm/year^{2}$. This result
might also suggest that the observed background multisecular acceleration
of about $0.01$ $mm/year^{2}$ has a natural origin and may be independent
of the late 20th century anthropogenic warming. Indeed, the quasi-millennial
solar/climate cycle observed throughout the Holocene has been in its
warming phase since the 17th century, which experienced the coldest
period of the Little Ice Age during the Maunder solar minimum, and
may reach a maximum during the second half of 21st century \citep{Humlum,Scafetta2012b}.

Using Eq. \ref{eq:1} for the 1700-2002 period, the regression rate
in $t_{0}=2000$ is $v_{0}=2.3\pm0.06$ $mm/year$. If the 1700-2002
acceleration, $a=0.0092\pm0.0004$ $mm/year^{2}$, persists during the
21th century, by extrapolating the 1700-2002 quadratic fit, the global
sea level may increase by $\sim277\pm7$ $mm$ from 2000 to 2100.

\begin{figure}[!t]
\includegraphics[width=1\columnwidth]{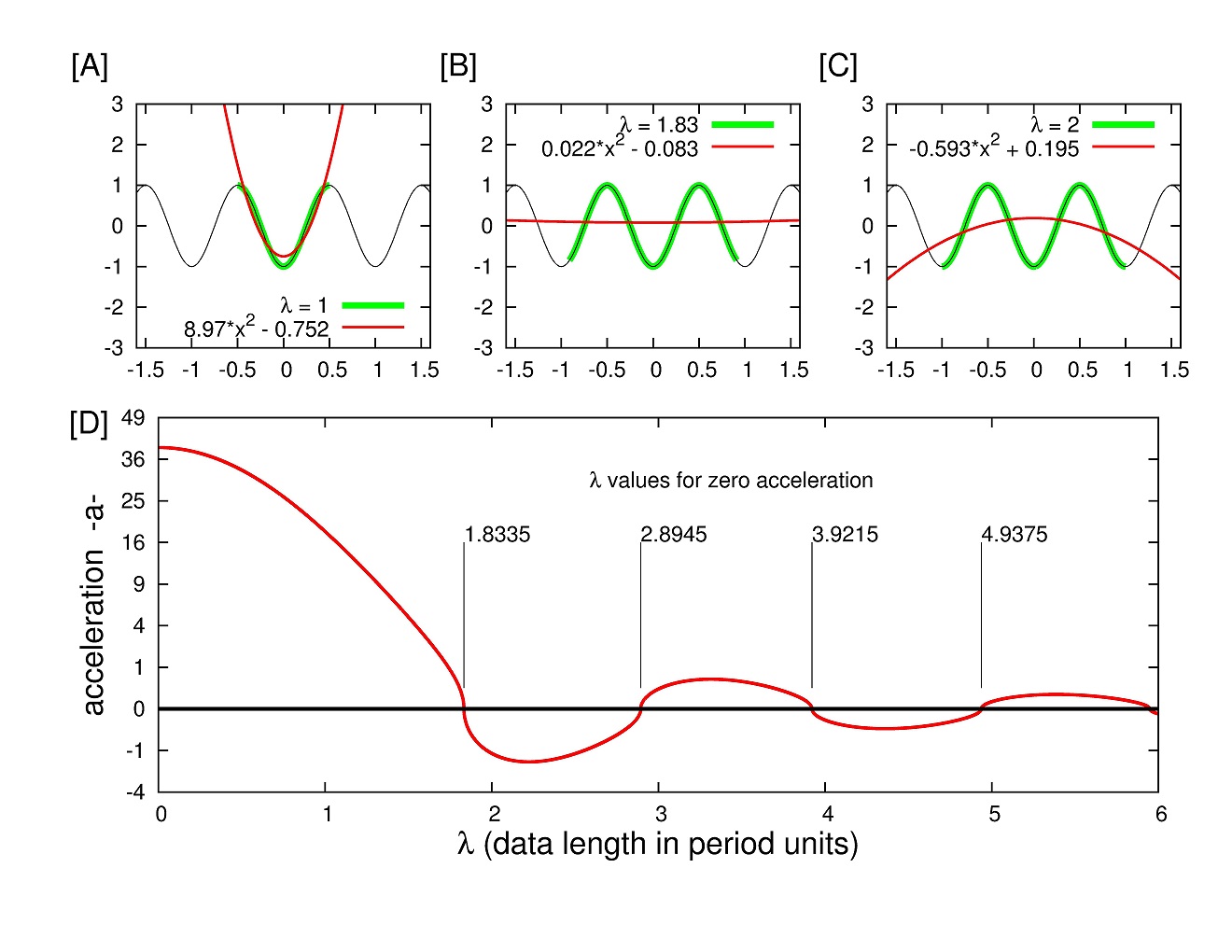}\caption{[A], [B] and [C] show a stationary harmonic signal (black) of unit period fit with Eq. \ref{eq:1} (red) within three different
intervals (indicated in green) of length $\lambda=$ $1$, $1.83$
and $2$, respectively. [D] The regression acceleration coefficient
in function of the record length $\lambda$. The figure demonstrates
that the acceleration oscillates in function of the length interval
used for the regression analysis. The values of $\lambda$ that make
the acceleration $a=0$, corresponding to a fully orthogonality between
the harmonic signal and the quadratic polynomial constructor are also
reported.}
\end{figure}

\section{Conclusion}

Herein I have proposed a multi-scale dynamical analysis (MSDA) to
study   climatic records. The proposed technique
uses graphical diagrams that greatly facilitate a comparative analysis
among tide gauge records and other climatic indexes such as PDO, AMO
and NAO. These records are found to be characterized by significant
oscillations at the decadal and multidcadal scales up to about 110-year
intervals. Within these scales both positive and negative accelerations
are found if a record is sufficiently long. This result suggests that
acceleration patterns in tide gauge records are mostly driven by the
natural oscillations of the climate system. The volatility of the
acceleration increases drastically at smaller scales such as at the
bi-decadal ones.

MSAA clearly reveals a macroscopic influence of the quasi 60-year
PDO and AMO oscillation in the tide gauge records, in particular for
Fremantle and New York City. In fact, although during the last 40-60 years
positive accelerations are typically observed in tide gauge records,
as found by \citet{Sallenger} and by \citet{Boon}, also during 40-60 year intervals since
the beginning of the 20th century equivalent or even higher positive
accelerations are typically observed. On the contrary, at these same
time scales, negative accelerations are observed during the middle
of the 20th century. This pattern demonstrates the influence of a
large multidecadal 60-year natural cycle, which has been found in
climate and solar/astronomical records by numerous authors \citep[e.g: ][]{Klyashtorin,Jevrejeva,Chambers,Scafetta2010,Scafetta2012m,Scafetta2013,Soon,Ogurtsov}.
The existence of a 60-70 year oscillation is evident in Figure 10
that shows the MSAA comparison between a global reconstruction of
the sea level and of the NAO index from 1700 to 2000. The figure demonstrates
that the sea level acceleration patterns during the last 60 years
are not particularly anomalous relative to previous periods because they
repeat quasi-periodically for three hundred years and well correlate
with an independent climatic index such as the long NAO reconstruction
record by \citet{Luterbacher,Luterbacher2002}.

Because of the presence of a quasi 60-year oscillation, a background
secular acceleration (e.g. one that could be caused by the 20th century
anthropogenic warming) can be properly evaluated only for tide gauge records
longer than about 110 years, as Figure 10A2 implicitly shows. If only
shorter periods are available, separating a background acceleration
from a 60-year oscillation is quite difficult because the record would
not be sufficiently long to determine the correct amplitude and phase
of the oscillation, which are required to separate the oscillation
from a background quadratic polynomial term. More precisely, if the
time interval to be fitted is significantly shorter than about 2 periods
of the oscillation, a quadratic polynomial constructor would be too
collinear with a sinusoidal oscillation, and in collinearity cases
the linear regression algorithm becomes highly unreliable. The test
depicted in Figure 11 demonstrates that the minimum length that a
record must have for making a quadratic polynomial constructor fully
orthogonal to a sinusoidal curve is $\lambda=1.8335$ the period of
the oscillation: that is, a 110-year long record is necessary for
fully filtering out a 60-year periodic cycle from a background quadratic polynomial
trend. In general, as Figure 11D implies, in the case of a record characterized
by a 60-year cycle, using 100-year or longer records would be sufficiently
fine, but using, for example, 60-year long or shorter records \citep[e.g. as done in ][]{Sallenger}
can be highly misleading because the quadratic fit would interpret
the bending of the 60-year oscillation as a strongly accelerating
trend.

When about 100-year long records are studied the evaluated average
accelerations are (positively or negatively) close to zero \citep{Church2011,Houston2011}.
Among the analyzed six tide gauge records (Sydney, Fremantle, New
York, Honolulu, San Diego and Venice) two records present a slight
positive secular acceleration ($\sim0.01$ $mm/year^{2}$), while the
other four records present a slight negative secular acceleration
($\sim-0.01$ $mm/year^{2}$). The 1700-2000 global sea level index
proposed by \citet{Jevrejeva} shows an almost uniform acceleration
($\sim0.01$ $mm/year^{2}$) at scales larger than 110 years spanning
from the Little Ice Age in 1700 to modern times. Thus, it is difficult
to determine whether the late 20th century anthropogenic warming had
a significant effect on tide gauge records. An anthropogenic warming
effect appears to be in any case too small to be clearly separated
from the multidecadal and multisecular natural variability. For example, because the global sea level
record proposed by \citet{Jevrejeva} presents an uniform secular-scale
acceleration of about $\sim0.01$ $mm/year^{2}$ since 1700, a theoretical
post-1900 sea level acceleration associated to the anthropogenic warming
could be constrained between $\sim0.0$ $mm/year^{2}$ and $\sim0.02$
$mm/year^{2}$, where the upper limit would apply in the eventuality
that the natural acceleration component from 1700 to 1900 was due
to a millennial oscillation that since 1900 changed inflection by
turning the natural component of the acceleration into a negative
value of $\sim-0.01$ $mm/year^{2}$. Indeed, the slight negative secular
acceleration found in a majority of tide gauge records \citep{Houston2011}
may also be due to the quasi millennial oscillation typically found
in numerous climatic and solar indexes throughout the Holocene, which
will reach its millennium maximum during the 21st century \citep{Humlum,Scafetta2012b}.

In conclusion, at scales shorter than 100-years, the measured tide
gauge accelerations are strongly driven by the natural oscillations
of the climate system (e.g. PDO, AMO and NAO). At the smaller scales
(e.g. at the decadal and bi-decadal scale) they are characterized
by a large volatility due to significant decadal and bi-decadal climatic
oscillations \citep{scafetta2009,Scafetta2010,Scafetta2012m,Manzi}.
Therefore, accelerations, as well as linear rates evaluated using
a few decades of data (e.g. during the last 20-60 years) cannot be
used for constructing reliable long-range projections of sea-level
for the 21st century. The oscillating natural patterns need to be included
in the models for producing reliable forecasts at multiple time scales.
The proposed MSDA methodologies (e.g. MSAA, MSRA and MSACAA) provide
a comprehensive picture to comparatively study  dynamical patterns in
tide gauge records. The techniques can be efficiently used for a quick
and robust study of alternative climatic sequences as well.

\newpage{}

\onecolumn

\section*{Appendix: data}
Data can be downloaded from the following web-sites:

Tide gauge records: \href{http://www.psmsl.org/data/obtaining/}{http://www.psmsl.org/data/obtaining/}

PDO: \href{http://jisao.washington.edu/pdo/PDO.latest}{http://jisao.washington.edu/pdo/PDO.latest}

AMO: \href{http://www.esrl.noaa.gov/psd/data/correlation/amon.us.long.data}{http://www.esrl.noaa.gov/psd/data/correlation/amon.us.long.data}

NAO: \href{http://www.esrl.noaa.gov/psd/gcos\_wgsp/Timeseries/RNAO/}{http://www.esrl.noaa.gov/psd/gcos\_wgsp/Timeseries/RNAO/}

Global sea level: \href{http://www.psmsl.org/products/reconstructions/jevrejevaetal2008.php}{http://www.psmsl.org/products/reconstructions/jevrejevaetal2008.php}


\begin{thebibliography}{}

\bibitem[Boon(2012)]{Boon} Boon, J.D., 2012. Evidence of sea level acceleration at U.S. and Canadian tide stations, Atlantic Coast, North America.
J. Coastal Research 28(6), 1437\textendash{}1445.

\bibitem[Boretti(2012)]{Boretti2012}Boretti, A., 2012. Is there any
support in the long term tide gauge data to the claims that parts
of Sydney will be swamped by rising sea levels? Coastal Engineering
64, 161\textendash{}167.

\bibitem[Chambers et al.(2012)]{Chambers} Chambers, D.P., Merrifield,
M.A., and Nerem, R.S., 2012. Is there a 60-year oscillation in global
mean sea level? Geophysical Research Letters 39, L18607.

\bibitem[Church and White(2006)]{Church2006}Church, J.A., White,
N.J., 2006. A 20th century acceleration in global sea level rise.
Geophysical Research Letters 33, L01602.

\bibitem[Church and White(2011)]{Church2011} Church, J.A., White,
N.J., 2011. Sea-level rise from the late 19th to the early 21st Century.
Surveys in Geophysics 32 (4\textendash{}5), 585\textendash{}602.

\bibitem[Chylek et al.(2011)]{Chylek}Chylek, P., Folland, C.K., Dijkstra,
H.A., Lesins, G., and Dubey, M.K., 2011. Ice-core data evidence for
a prominent near 20 year time-scale of the Atlantic Multidecadal Oscillation.
Geophysical Research Letters 38, L13704.

\bibitem[Douglas(1992)]{Douglas1992}Douglas, B. C., 1992. Global
sea level acceleration. J. Geophys. Res. 97, 12699-12706.

\bibitem[Dean and Houston(2013)]{Dean2013} Dean, R.G., Houston, J.R.,
2013. Recent sea level trends and accelerations: Comparison of tide
gauge and satellite results. Coastal Engineering, 75, 4\textendash{}9.

\bibitem[Houston and Dean(2011)]{Houston2011} Houston, J.R., Dean,
R.G., 2011. Sea-Level Acceleration Based on U.S. Tide Gauges and Extensions
of Previous Global-Gauge Analyses. Journal of Coastal Research 27,
409\textendash{}417.

\bibitem[Humlum et al.(2011)]{Humlum}Humlum, O., Solheim, J.-K.,
Stordahl, K., 2011. Identifying natural contributions to late Holocene
climate change. Global and Planetary Change 79, 145\textendash{}156.

\bibitem[Hunter and Brown(2013)]{Hunter}Hunter, J.R., Brown, M.J.I,
2013. Discussion of Boretti, A., \textquoteleft{}Is there any support
in the long term tide gauge data to the claims that parts of Sydney
will be swamped by rising sea levels?\textquoteright{}, Coastal Engineering,
64, 161\textendash{}167, June 2012. Coastal Engineering 75, 1\textendash{}3.

\bibitem[IPCC(2007)]{IPCC2007} IPCC: Solomon, S., et al. (eds) in
Climate Change 2007: The Physical Science Basis.Contribution of Working
Group I to the Fourth Assessment Report of the Intergovernmental Panel
on Climate Change, (Cambridge University Press, Cambridge, 2007).

\bibitem[Jevrejeva et al.(2008)]{Jevrejeva}Jevrejeva, S., et al.,
2008. Recent global sea level acceleration started over 200 years
ago? Geophysical Research Letters 35, L08715.

\bibitem[Kennedy et al.(2011a)]{Kennedy} Kennedy, J. J., Rayner,
N. A., Smith, R. O., Saunby, M., Parker, D. E., 2011a. Reassessing
biases and other uncertainties in sea-surface temperature observations
since 1850 part 1: measurement and sampling errors. J. Geophys. Res.,
116, D14103.

\bibitem[Kennedy et al.(2011b)]{Kennedyb} Kennedy, J. J., Rayner,
N. A., Smith, R. O., Saunby, M. and Parker, D. E., 2011b. Reassessing
biases and other uncertainties in sea-surface temperature observations
since 1850 part 2: biases and homogenisation. J. Geophys. Res., 116,
D14104, doi:10.1029/2010JD015220

\bibitem[Klyashtorin et al.(2009)]{Klyashtorin}Klyashtorin, L.B.,
Borisov, V., Lyubushin, A., 2009. Cyclic changes of climate and major
commercial stocks of the Barents Sea. Marine Biology Research 5, 4-17.

\bibitem[Knudsen et al.(2011)]{Knudsen}Knudsen, M.F., Seidenkrantz,
M-S., Jacobsen, B.H., Kuijpers, A., 2011. Tracking the Atlantic Multidecadal
Oscillation through the last 8,000 years. Nature Communications 2,
178.

\bibitem[Kobashi et al.(2010)]{Kobashi}Kobashi, T., Severinghaus,
J., Barnola, J.-M., Kawamura, K., Carter, T., Nakaegawa, T., 2010.
Persistent multi-decadal Greenland temperature fluctuation through
the last millennium. Climate Change 100, 733-756.

\bibitem[Levermann et al.(2005)]{Levermann}Levermann, A., Griesel,
A., Hofmann, M., Montoya, M., Rahmstorf, S., 2005. Dynamic sea level
changes following changes in the thermohaline circulation. Clim. Dynam.
24, 347\textendash{}354.

\bibitem[Loehle and Scafetta(2011)]{Loehle}Loehle, C., N. Scafetta,
N., 2011. Climate Change Attribution Using Empirical Decomposition
of Climatic Data. The Open Atmospheric Science Journal 5, 74-86.

\bibitem[Luterbacher et al.(1999)]{Luterbacher}Luterbacher, J., Schmutz,
C., Gyalistras, D., Xoplaki, E., Wanner, H., 1999. Reconstruction
of monthly NAO and EU indices back to AD 1675. Geophys. Res. Lett.
26, 2745-2748.

\bibitem[Luterbacher et al.(2002)]{Luterbacher2002} Luterbacher,
J., Xoplaki, E., Dietrich, D., Jones, P. D., Davies, T. D., Portis,
D., Gonzalez-Rouco, J. F., von Storch, H., Gyalistras, D., Casty,
C., and Wanner, H., 2002. Extending North Atlantic Oscillation Reconstructions
Back to 1500. Atmos. Sci. Lett., doi:10.1006/asle.2001.0044.

\bibitem[Mantua et al.(1997)]{Mantua}Mantua, N. J., Hare, S. R.,
Zhang, Y., Wallace, J. M., Francis, R. C., 1997. A Pacific Interdecadal
Climate Oscillation with Impacts on Salmon Production. Bull. Amer.
Meteor. Soc. 78, 1069\textendash{}1079.

\bibitem[Manzi et al.(2012)]{Manzi}Manzi V., Gennari, R., Lugli,
S., Roveri, M., Scafetta, N., Schreiber, C., 2012. High-frequency
cyclicity in the Mediterranean Messinian evaporites: evidence for
solar-lunar climate forcing. Journal of Sedimentary Research 82, 991-1005.

\bibitem[Mazzarella and Scafetta(2012)]{Mazzarella}Mazzarella, A.,
Scafetta, N., 2012. Evidences for a quasi 60-year North Atlantic Oscillation
since 1700 and its meaning for global climate change. Theoretical
and Applied Climatology 107, 599-609.

\bibitem[M\"orner(1989)]{Morner1989}M\"orner, N.-A., 1989. Changes
in the Earth's rate of rotation on an El Nino to century basis. In:
Geomagnetism and Paleomagnetism ( F.J. Lowes et al., eds), 45-53,
Kluwer Acad. Publ.

\bibitem[M\"orner(1990)]{Morner(1990)}M\"orner, N.-A., 1990. The
Earth's differential rotation: hydrospheric changes. Geophysical Monographs,
59, 27-32, AGU and IUGG.

\bibitem[M\"orner(2010)]{Morner}M\"orner, N.-A., 2010. Some problems
in the reconstruction of mean sea level and its changes with time.
Quaternary International 221 (1-2), 3-8.

\bibitem[Ogurtsov et al.(2002)]{Ogurtsov}Ogurtsov, M. G., Nagovitsyn,
Y. A., Kocharov, G. E., Jungner, H., 2002. Long-period cycles of the
Sun\textquoteright{}s activity recorded in direct solar data and proxies.
Solar Physics 211, 371-394.

\bibitem[Parker(2012a)]{Parker2012a}Parker, A., 2012a. Oscillations
of sea level rise along the Atlantic coast of North America north
of Cape Hatteras. Natural Hazards 65, 991\textendash{}997.

\bibitem[Parker(2012b)]{Parker2012b}Parker, A., 2012b. Sea level
trends at locations of the United States with more than 100 years
of recording. Natural Hazards 65, 1011\textendash{}1021.

\bibitem[Parker(2013)]{Parker}Parker, A., 2013. Natural oscillations
and trends in long-term tide gauge records from the Pacific. Pattern
Recogn. Phys., 1, 1-13.

\bibitem[PSMLS(2013)]{PSMSL2013} Permanent Service for Mean Sea Level
(PSMSL), 2013. Tide Gauge Data. The records were downloaded during
Feb 2013.

\bibitem[Sallenger et al.(2012)]{Sallenger} Sallenger Jr., A.H.,
Doran, K.S., Howd, P.A., 2012. Hotspot of accelerated sea-level rise
on the Atlantic coast of North America. Nature Climate Change 2, 884-888.

\bibitem[Scafetta(2009)]{scafetta2009}Scafetta N., 2009. Empirical
analysis of the solar contribution to global mean air surface temperature
change. Journal of Atmospheric and Solar-Terrestrial Physics 71, 1916-1923.

\bibitem[Scafetta(2010)]{Scafetta2010}Scafetta, N., 2010. Empirical
evidence for a celestial origin of the climate oscillations and its
implications. Journal of Atmospheric and Solar-Terrestrial Physics
72, 951-970.

\bibitem[Scafetta(2012a)]{Scafetta2012m}Scafetta N., 2012a. Testing
an astronomically based decadal-scale empirical harmonic climate model
versus the IPCC (2007) general circulation climate models. Journal
of Atmospheric and Solar-Terrestrial Physics 80, 124-137.

\bibitem[Scafetta(2012b)]{Scafetta2012b}Scafetta N., 2012b. Multi-scale
harmonic model for solar and climate cyclical variation throughout
the Holocene based on Jupiter-Saturn tidal frequencies plus the 11-year
solar dynamo cycle. Journal of Atmospheric and Solar-Terrestrial Physics
80, 296-311.

\bibitem[Scafetta and Willson(2013)]{Scafetta2013}Scafetta, N., Willson,
R. C., 2013. Planetary harmonics in the historical Hungarian aurora
record (1523\textendash{}1960). Planetary and Space Science 78, 38-44.

\bibitem[Schlesinger and Ramankutty(1994)]{Schlesinger}Schlesinger,
M. E., Ramankutty, N., 1994. An oscillation in the global climate
system of period 65-70 years. Nature 367, 723-726.

\bibitem[Soon and Legates(2013)]{Soon}Soon, W., Legates, D. R., 2013.
Solar irradiance modulation of Equator-to-Pole (Arctic) temperature
gradients: Empirical evidence for climate variation on multi-decadal
timescales. J. of Atmospheric and Solar-Terrestrial Physics 93, 45-56.

\bibitem[Woodworth(1990)]{Woodworth1990} Woodworth, P. L., 1990.
A search for accelerations in records of European mean sea-level.
Int. J. Climatol. 10, 129-143.

\bibitem[Woodworth and Player(2003)]{Woodworth(2003)} Woodworth,
P. L., Player, R., 2003. The Permanent Service for Mean Sea Level:
an update to the 21st century. Journal of Coastal Research, 19, 287-295.

\bibitem[Woodworth et al.(2009)]{Woodworth2009}Woodworth, P. L.,
White, N. J. Jevrejeva, S., Holgate, S. J., Church, J. A., Gehrels,
W. R., 2009. Evidence for the accelerations of sea level on multi-decade
and century timescales. Int. J. Climatol. 29, 777-789.

\end{thebibliography}
\end{document}